\documentclass[prd,aps,floatfix,superscriptaddress,twocolumn, nofootinbib]{revtex4-2}

\usepackage{enumerate}
\usepackage{bm}
\usepackage{bbm}
\usepackage[usenames,dvipsnames]{xcolor}
\usepackage{amssymb,amsmath}
\usepackage{graphicx}
\usepackage{overpic}
\usepackage{color}
\usepackage[colorlinks=true,citecolor=magenta]{hyperref}
\usepackage{enumitem}
\usepackage{braket}
\usepackage{tensor}
\usepackage{physics}

\def\ba{\begin{eqnarray}}
\def\ea{\end{eqnarray}}
\newcommand{\mpl}{M_{\rm{Pl}}}
\newcommand{\cut}{\Lambda}

\def\({\left(}
\def\){\right)}
\def\L{\mathcal{L}}

\begin{document}

\title{Probing higher-spin particles with gravitational waves from compact binary inspirals}

\author{Hao-Yang Liu}
\email{liuhaoyang19@mails.ucas.ac.cn}
\affiliation{International Centre for Theoretical Physics Asia-Pacific, University of Chinese Academy of Sciences, 100190 Beijing, China}
\affiliation{School of Physics, University of Chinese Academy of Sciences, Beijing 100049, China}

\author{Yun-Song Piao}
\email{yspiao@ucas.ac.cn}
\affiliation{School of Physics, University of Chinese Academy of Sciences, Beijing 100049, China}
\affiliation{International Centre for Theoretical Physics Asia-Pacific, University of Chinese Academy of Sciences, 100190 Beijing, China}
\affiliation{School of Fundamental Physics and Mathematical Sciences, Hangzhou Institute for Advanced Study, UCAS, Hangzhou 310024, China}
\affiliation{Institute of Theoretical Physics, Chinese Academy of Sciences, P.O. Box 2735, Beijing 100190, China}

\author{Jun Zhang}
\email{zhangjun@ucas.ac.cn}
\affiliation{International Centre for Theoretical Physics Asia-Pacific, University of Chinese Academy of Sciences, 100190 Beijing, China}
\affiliation{Taiji Laboratory for Gravitational Wave Universe (Beijing/Hangzhou), University of Chinese Academy of Sciences, 100049 Beijing, China}

\begin{abstract}
Under the framework of gravitational effective field theory, we propose a theory agnostic strategy of searching for higher-spin particles with gravitational waves from compact binary inspirals. Using this strategy, we analyze gravitational wave signals from the binary black hole merger events GW151226 and GW170608, as well as the binary neutron star merger event GW170817. We find that the existence of higher-spin particles with mass ranged from $10^{-12} {\rm eV}$ to  $10^{-11} {\rm eV}$ is strongly disfavored by these events unless the particles precisely combine within a supersymmetric supermultiplet. We argue that the gravitational effective field theory also provides a framework to search for signals beyond GR from other GW observations, such as extreme-mass-ratio-inspirals.
\end{abstract}

\maketitle

\textbf{Introduction.---}
Particles beyond the standard model are generally expected in our universe.
Their existence is not only predicted by potential UV completion theories such as string theory, but also indicated by the observations of dark matter and dark energy. Searching for such particles, however, can be very challenging, as these particles could be dark --- in the sense that they only weakly couple to or even do not directly interact with the stand model particles. While most models of dark matter and dark energy have focused on the presence of weak couplings between the dark sector and baryonic matter, the only guaranteed coupling with this dark sector is gravitational. The imprint of the dark sector within the effective field theory (EFT) of gravity should therefore be understood as a privileged and model independent way to probe the dark sector. 

Naturally, when only accounting for gravitational interactions, the imprint of dark sector is expected to be Planck scale suppressed, however cosmological probes and gravitational wave (GW) astronomy has reached such a level of precision that we entering the realm where an ultralight dark sector would have a small yet detectable effect on the waveform.
Particularly, given the rapid development in GW astronomy, we are able to test General Relativity (GR) with high precision~\cite{LIGOScientific:2021sio}, and even to hunt for dark particles, for example, through their direct emission of GWs~\cite{Tsukada:2018mbp, Palomba:2019vxe, Sun:2019mqb, Tsukada:2020lgt, KAGRA:2021tse, Yuan:2022bem} and their imprints on the coalescence waveform~\cite{Zhang:2021mks, Maselli:2021men, Barsanti:2022ana, Zhang:2022rfr, Barsanti:2022vvl}. One subtlety of probing the dark sector with GWs is to characterized the effects from the dark particles on GWs while the exact theory of the dark particles remains unknown. This is especially the case for the match-filtering analysis of GW signals, where the waveform template is needed a priori. Current theory agnostic searches mostly rely on parameterized tests, where the waveform is constructed by promoting the coefficients in the post-Newtonian inspiral waveform to unknown parameters, so that the waveform could capture some possible deviations from GR \cite{LIGOScientific:2021sio, Yunes:2009ke, Yunes:2016jcc}.

In this letter, we propose a theory agnostic strategy of searching for the dark particles under the framework of gravitational EFT \cite{Donoghue:1995cz, Donoghue:1994dn}. If one only focuses on the gravitational phenomena at energy much lower than the mass of dark particles, one can integrate out the dark particles and obtain an EFT. In this case, the effects of the dark particles on the low energy gravity are captured by higher dimension EFT operators, and the Wilson coefficients of the EFT are closely related to the properties of the dark particles. In practice, taking a bottom up approach, one can construct a general EFT that includes all possible dominating effects from heavy particles, and use it as a guide for potential hints of dark particles. We analyze GW signals from three binary merger events using this strategy, and report the first constraints on higher-spin particles from GW observations.\\

\textbf{Gravitational EFT.---}
We shall assume Lorentz invariance and parity conservation, and consider the most general gravitational EFT that propagates two massless spin-2 degrees of freedom, the action of which can be arranged by the dimension of the operators as follow \cite{Donoghue:1995cz, Donoghue:1994dn, Ruhdorfer:2019qmk}:
\ba
\label{eq:EFT}
{\cal L}_{\rm EFT} = \frac{\mpl^2}{2} \left( R  + \frac{{\cal L}_{\rm D4}}{\Lambda_4^2} + \frac{{\cal L}_{\rm D6}}{\cut_6^4} + \frac{{\cal L}_{\rm D8}}{\cut_8^6} + \cdots
 \right)\,,
\ea
where ${\cal L}_{{\rm D}n}$ denotes a linear combination of all possible dim-$n$ operators built out of the Riemann (or Weyl) curvature tensor and its covariant derivatives, and is suppressed by a scale $\Lambda_n$. As stated previously, the higher dimension operators are obtained by integrating out the dark particles when focusing on their effects on the low energy gravity. In particular, these operators could come from integrating out $N$ massive particles of mass $m$ and spin $\le 2$ at loop level, in which case, dimension-$n$ operators are then suppressed by the scale $\Lambda_n=\(m^{n-4}\mpl^2/N\)^{1/(n-2)}$~ \cite{Avramidi:1986mj, deRham:2019ctd}. Bearing in mind that the species bound $Nm^2\lesssim \mpl^2$ should be satisfied, this provides an upper bound on the number $N$ of particles at the mass $m$ and in practise loop effects are only competitive with tree-level ones when considering sufficiently high order operators $n\ll 1$ which are typically irrelevant.
Alternatively, one can also consider these operators coming from integrating out $N$ massive particles of mass $M$ and spin $\ge 3$ at tree level, in which case the cutoff is given by $\Lambda_n=M/N^{1/(n-2)}$ (see Ref.~\cite{deRham:2019ctd} for further discussions). In what follows we shall consider the latter situation and assume, without much loss of generality, that $N\sim \mathcal{O}(1)$. In this case, we do not distinguish $\Lambda_n$ and define the cut-off scale $\Lambda \simeq \Lambda_n$. In fact, precisely how the number of particles affect the low-energy EFT depends on its details. For a supersymmetric dark sector, the contribution to the dimension-6 operators would precisely cancel. In what follows we shall therefore consider the imprint of higher-spin bearing in mind that we expect supersymmetry to be broken at the scales we are interested in. It is worth noting that  the presence of higher dimensional operators $\L_{\rm D8}$ would also be linked with the presence of higher-spin particles. Generic constraints from high energy completions were considered in Refs.~\cite{Bern:2021ppb,Caron-Huot:2022ugt,deRham:2022gfe}. Unless the specific assumption of a perfectly supersymmetric dark sector is made,  the effect of dimension-6 are always expected to dominate over the more irrelevant dimension-8 operators in phenomenologically relevant situations.\\

\textbf{Inspiral waveform.---}
Black holes and GWs in gravitational EFTs have been studied in Refs.~\cite{Endlich:2017tqa, Cardoso:2018ptl, deRham:2020ejn, Cano:2021myl}. It is understood that, for Ricci-flat solutions, the dim-4 operators do not contribute, and the leading EFT corrections come from the dim-6 operators. Moreover, among all possible dim-6 operators, only two operators contribute to Ricci flat solutions independently,\footnote{In principle, one can remove one of the two operators by field redefinition, which, however, will introduce additional couplings in the matter sector. In this letter, we shall work in the frame where both operators are present. Interestingly, only one combination of those is constrained by current positivity and causality bounds \cite{Bern:2021ppb,Caron-Huot:2022ugt,deRham:2022gfe}, while the other remains so far unconstrained.} and shall be considered in binary inspirals. Without loss of generality, we write
\ba
{\cal L_{\rm D6}} = \alpha_1  I_1 + \alpha_2 \left(I_1 - 2 I_2\right) + {\cal L_{\rm D6}^{\rm tri}}  ,
\ea
where~$I_1$ and $I_2$ are the two independent dim-6 operators $ \tensor{R}{_\mu_\nu^\alpha^\beta} \tensor{R}{_\alpha_\beta^\gamma^\sigma} \tensor{R}{_\gamma_\sigma^\mu^\nu}$ and $ \tensor{R}{_\mu^\alpha_\nu^\beta}\tensor{R}{_\alpha^\gamma_\beta^\sigma} \tensor{R}{_\gamma^\mu_\sigma^\nu}$, and ${\cal L_{\rm D6}^{\rm tri}}$ denotes the other dim-6 operators that do not contribute. $\alpha_1$ and $\alpha_2$ are the Wilson coefficients, which are expected to be of order of unity.

Notably, not all of the lower energy EFTs, namely model \eqref{eq:EFT} with arbitrary coefficients, can be embedded into a sensible UV complete theory. For example, it has been shown that the possible values of $\alpha_1$ and $\alpha_2$ are constrained, together with coefficients dim-8 operators, from positive bounds \cite{Caron-Huot:2022ugt}. By considering GWs scattering on a Schwarzschild-like black hole, infrared causality also demands $\alpha_1$ to be nonnegative, assuming the higher dimension operators coming from tree level interactions of higher-spin particles \cite{deRham:2021bll, deRham:2022hpx}.  For lower spin particles, explicit calculation shows that $\alpha_1$ is positive if the particles are Bosons, and is negative if the particles are Ferimions \cite{Avramidi:1986mj, deRham:2019ctd, Bern:2021ppb, Caron-Huot:2022ugt}. Nevertheless, we shall not consider negative $\alpha_1$, because the observed GW signals generally beyond the EFT validity regime if the higher dimension operators come from lower spin particles. These theoretical constraints can serve as prior in the later Bayesian analysis, where we set $\alpha_1 = 1$ for a given $\Lambda$, and consider $-10 < \alpha_2 < 10$ as $\alpha_2$ is expected to be of ${\cal O}(1)$.

Corrections from the dim-6 operators on the gravitational potential and the power of GW radiation of a binary system have been studied in Refs.~\cite{Brandhuber:2019qpg, Emond:2019crr, AccettulliHuber:2020dal}, given which one can obtain the corresponding corrections on the inspiral waveform~\cite{AccettulliHuber:2020dal}. Under the stationary phase approximation, the inspiral waveform can be constructed in the frequency domain: 
\begin{align}
h(f)_{\rm  SPA} \simeq H(f)\ e^{i \Psi(f)} \, .
\end{align}
Focusing on the phase of the waveform, $\Psi(f)$ can be written into three parts,
\ba
\Psi(f) = \Psi_{\rm GR}(f) + \Psi_{\rm tidal}(f) + \Psi_{\rm D6}(f) \, .
\ea
Here $\Psi_{\rm GR}(f)$ is the phase obtained in GR, and is given by, e.g., the usual TaylorF2 template~\cite{lalsuite}. $\Psi_{\rm tidal}(f)$ represents the phase evolution caused by tidal deformation of the compact objects in the binary, and shall be turned on only for binary neutron star inspirals. $\Psi_{\rm D6}(f)$ is the corrections from the dim-6 operators. To the leading order in post-Newtonian (PN) expansion, we have~\cite{AccettulliHuber:2020dal},
\begin{align}\label{eq:psiD6}
\Psi_{\rm D6} = &- \frac{3}{128 \nu v_f^5} \bigg[ \frac{936\alpha_2 }{(GM\Lambda)^4}  v_f^{10} \nonumber \\
&-\frac{13080 \alpha_1+(3990-5100 \nu)  \alpha_2}{7(GM\Lambda)^4}v_f^{12}\bigg],
\end{align}
where $M$ is the total mass of binary, $\nu$ is the dimensionless reduced mass, and $v_f\equiv(\pi GM f)^{1/3}$.

Note that Eq.~\eqref{eq:psiD6} and hence the corrected inspiral waveform are only justified within the validity regime of the EFT. To assure the validity of the EFT, we require the size of the binary to be much larger than $1/\Lambda$, i.e., the length scale corresponding to the EFT cut-off. In terms of the GW frequency, this requirement translates into
\ba
f \ll f_{\Lambda} \equiv \sqrt{GM\Lambda^3} \, .
\ea
In practice, we define $f_{\rm cut} = f_\Lambda/4$ and assume Eq.~\eqref{eq:psiD6} is valid as long as $f \le f_{\rm cut}$. When analyzing GW data, we further define $f_{\rm max}$ to be the GW frequency of innermost stable circular orbit (ISCO) or $f_{\rm cut}$, whichever is smaller, and only use the data with frequency lower than $f_{\rm max}$.

Moreover, for binary neutron star inspirals, we assume the two neutron stars obey the same equation of state (EOS), and hence their tidal deformabilities $\Lambda_1$ and $\Lambda_2$ are related. Following Ref.~\cite{Abbott:2018exr}, we consider that the symmetric tidal deformability $\Lambda_s \equiv (\Lambda_2+\Lambda_1)/2$, the antisymmetric tidal deformability $\Lambda_a \equiv (\Lambda_2-\Lambda_1)/2$ and the mass ratio of the binary $q \equiv m_2/m_1 \le 1$ are related through an EOS-insensitive relation $\Lambda_a(\Lambda_s, q)$ \cite{Yagi:2013bca,Yagi:2013awa}. To take into account uncertainties from the EOS, the relation $\Lambda_a(\Lambda_s, q)$ is tuned to a large set of EOS models \cite{Yagi:2015pkc,Chatziioannou:2018vzf}. When performing Bayesian analysis, we sample in the symmetric tidal deformability $\Lambda_s$, while $\Lambda_a$ and hence $\Lambda_1$ and $\Lambda_2$ are obtained using the EOS-insensitive relation $\Lambda_a(\Lambda_s, q)$.\\

\begin{figure}[tp]
\centering
\includegraphics[height=0.3\textwidth]{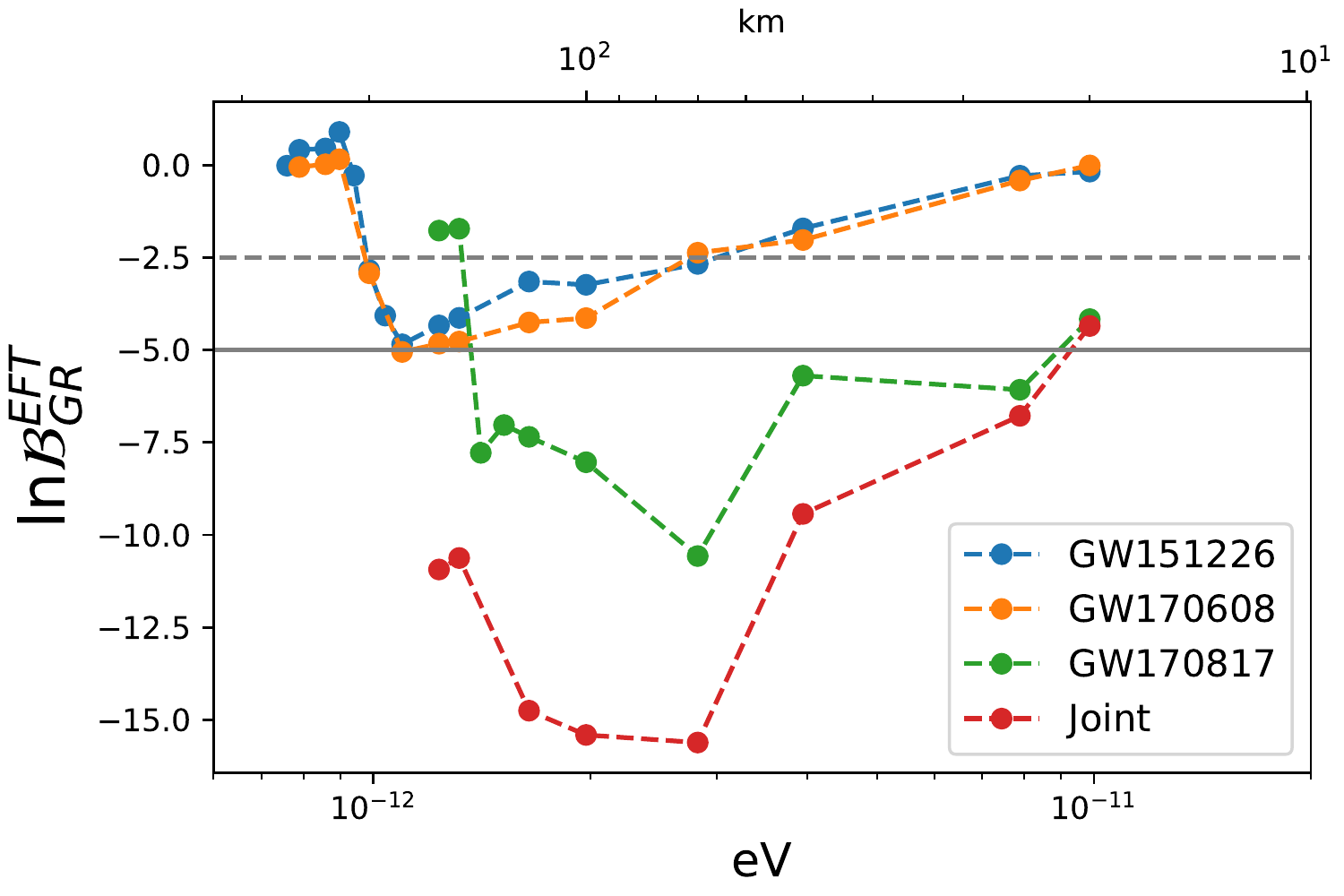}
\caption{Bayes factors comparing GR and EFT of different cut-off scale $\Lambda$ given the two binary black hole merger events and the binary neutron star merger event. The EFT is moderately disfavored if ${\cal B}^{\rm EFT}_{\rm GR} < 10^{-2.5}$ and is strongly disfavored if ${\cal B}^{\rm EFT}_{\rm GR} < 10^{-5}$.}\label{fig:lnBF}
\end{figure}

\textbf{Bayesian analysis.---} In order to search for the EFT corrections, we scan the parameter space by considering gravitational EFT \eqref{eq:EFT} with different cut-off scales. For each cut-off scale $\Lambda$, we perform Bayesian analysis, comparing the hypothesis that the observed GW signals are well described by the EFT and the hypothesis that the signals are well described by GR.

Specifically, we assume a Gaussian noise model, and define the likelihood function to be the distribution of the residuals. We also assume a minimum frequency $f_{\rm min}$, and only analyze GW data with frequency between $f_{\rm min}$ and $f_{\rm max}$ when computing likelihood function. Then we can compute the evidence for both hypothesis,
\ba
p\left(d|{\cal H}, I\right) = \int p\left( \bm{\theta} | {\cal H}, I\right) p\left(d| \bm{\theta}, {\cal H}, I\right) {\rm d} \bm{\theta}\, ,
\ea
where ${\cal H}$ can be $\rm EFT$ or $\rm GR$, denoting the hypothesis, $\bm \theta$ is the parameters in the waveform, $d$ is the observed GW data, and $I$ denotes the prior. In practice, the evidence and the posterior density functions of $\bm \theta$ are obtained with a nested sampling algorithm as implemented in the \verb"parallel bilby" package~\cite{Smith:2019ucc, Ashton:2018jfp, Speagle:2019ivv, skilling2004, skilling2006}. Given the evidence, we can compare the two hypothesis with the Bayes factor given the observed GW data $d$,
\ba
\left. {\cal B}^{\rm EFT}_{\rm GR} \right|_{d} = \frac{  p\left(d|{\rm EFT}, I\right)}{p\left(d|{\rm GR},I\right)}\, .
\ea

For GW data, we consider two binary black hole merger events GW151226 and GW170608, which have relatively long inspiral signals, and take $f_{\rm min}$ to be $20 {\rm Hz}$ and $30 {\rm Hz}$ respectively. When assuming GR, the sampled parameters $\bm \theta$ includes chirp mass ${\cal M}$, mass ratio $q$, coalescence time, coalescence phase, polarization, inclination, spins of two BHs, luminosity distance and the sky location. The prior probability density $P({\bm \theta}|{\rm GR}, I)$ is chosen as in Ref.~\cite{Veitch:2014wba}. When assuming EFT, we have an extra parameter $\alpha_2$ in the parameter set $\bm\theta$. For the prior $P({\bm \theta}|{\rm EFT}, I)$, we choose $\alpha_2$ to be uniformly distribute between $-10$ and $10$, while the prior of the other parameters are the same as in GR. We also analyze the binary neutron star merger event GW170817 with $f_{\rm min} = 23 {\rm Hz}$. In this case, the waveform template involves an additional parameter $\Lambda_s$, which corresponds to the tidal deformability of the NSs, and is uniformly sampled in the range of $[0,\,5000]$.

\begin{figure}[tp]
\centering
\includegraphics[height=0.275\textwidth]{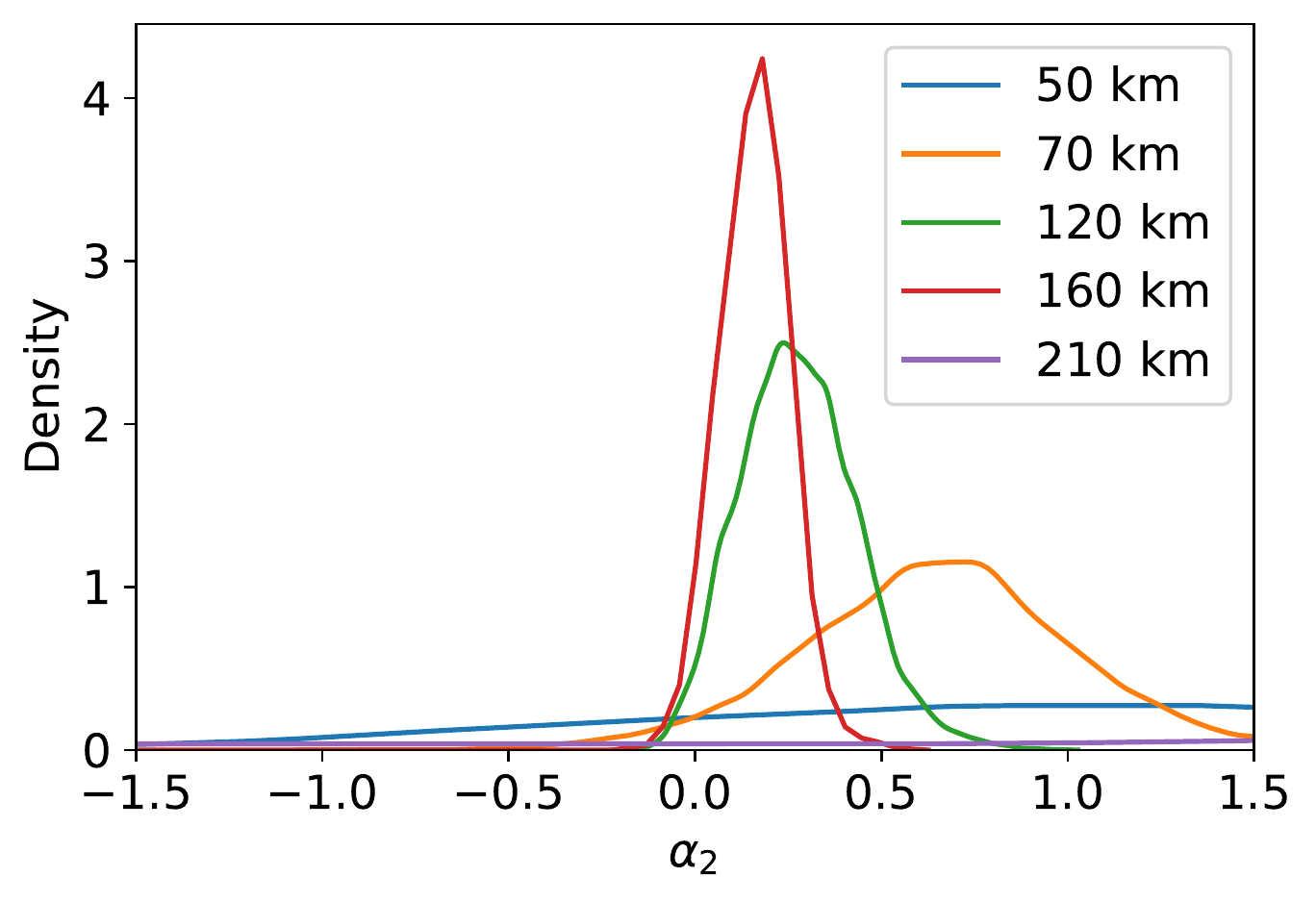}
\caption{Posterior of $\alpha_2$ given the binary black hole merger event GW151226. In this plot, we shows the posterior of $\alpha_2$ for $1/\Lambda = 50,\,70,\,120,\,160,\, 210\,{\rm km}$ in blue, orange, green, red and purple.}\label{fig:a2}
\end{figure}

The resulting Bayes factors for different $\Lambda$ are shown in Fig.~\ref{fig:lnBF}. Given three events, we can also define a joint Bayes factor,
\ba
\bar{\cal B}^{\rm EFT}_{\rm GR} = \prod_{i=1}^{3}  \left. \cal B^{\rm EFT}_{\rm GR}\right|_{d_i} \, .
\ea
In addition, the posterior of $\alpha_2$ is shown in Fig~\ref{fig:a2}.\\

\textbf{Implications on the dark particles.---} According to the Bayes' theorem, the EFT is preferred by the observations over GR, if the Bayes factor is much larger than $1$. From Fig.~\ref{fig:lnBF}, we find a moderate disfavor range for the cut-off scale, $\Lambda \in \left[9.87\times 10^{-13}, 2.82\times 10^{-12}\right] {\rm eV}$, given the two binary black hole merger events, and a strong disfavor range, $\Lambda \in \left[1.41\times 10^{-12}, 7.89\times 10^{-12}\right] {\rm eV}$, given the binary neutron star merger event. Joining the three events, we can exclude the EFT with $\Lambda \in \left[10^{-12}, 10^{-11}\right] {\rm eV}$.

Beyond this range, current GW observations lose the capability of distinguishing EFT from GR: As $\Lambda$ approaches to $ 0.97\times 10^{-13} {\rm eV}$, we have $f_{\rm cut} \sim 40 {\rm Hz}$, leaving no much data available for analysis and for distinguishing the two hypothesis. This is especially the case for the binary neutron star event, where the inspiraling objects have lower masses, so that one cannot distinguish EFT from GR when $\Lambda$ towards $ 1.32\times 10^{-12} {\rm eV}$. As $\Lambda$ approaches to $9.87\times 10^{-12} {\rm eV}$, we have $f_{\rm cut} \sim 1400 {\rm Hz}$ for binary black hole merger events, and we can use the GW data up to the ISCO frequency. The phase shift caused by the EFT correction, however, is about $ \mathcal{O}(0.1) $, making the EFT undistinguishable from GR.

Constraints on the cut-off scale can be directly casted on the mass of the higher-spin particles as $m \sim \Lambda$. Namely, the existence of higher-spin particles with mass ranged in $\left[10^{-12}, 10^{-11}\right] {\rm eV}$ is strongly disfavored by these three GW events. Our results are, however, cannot impose constraints on the lower spin particles. Because their mass $m \sim \Lambda^2 N^{1/2}/\mpl$, and is much lower than $\Lambda$ for what we have considered, even for a reasonably large $N$~\cite{Caron-Huot:2022ugt}. In this case, the EFT corrections are highly suppressed and hence are hardly detectable if we require the size of the binary remains larger than $1/m$ so that the inspiral waveform is justified.
\\

\textbf{Discussion.---}In our analysis, we require $\alpha_1$ to be positive given the theoretical constraints from positivity bounds and infrared causality. Actually, theoretical considerations also indicate $\alpha_2 > -\Lambda^2/\mpl$ \cite{deRham:2021bll}, which effectively means $\alpha_2>0$ for the $\Lambda$ that we considered, and could be used as prior in our analysis. We choose the prior to be $-10 < \alpha_2 < 10$ in our analysis, and we argue that requiring $\alpha_2$ to be positive should not change our results, as the posterior mostly distributes in the regime of $\alpha_2 > 0$, c.f. Fig.~\ref{fig:a2}.

We do not consider corrections from dim-8 and even higher dimension operators as those are irrelevant as compared to dim-6 ones, even in the regime we work in, where $GM\Lambda \gg 1$. The phase corrections from the dim-8 operators $\Psi_{\rm D8}$ have been studied in Ref.~\cite{Endlich:2017tqa}. Comparing to the dim-6 operators, we have $\Psi_{\rm D8}/\Psi_{\rm D6} \sim v_f^4/(GM\Lambda)^2 \sim (f_{\rm min}/f_\Lambda)^{4/3} \ll 1$. Therefore, the dim-8 operators can be safely neglected in our analysis. 
While dim-6 would be absent in perfectly supersymmetric realizations, supersymmetry is necessarily broken at the scales we are interested in and it would therefore be unnatural to switch the dim-6 operators off while maintaining dim-8 or higher order ones. The connection between irrelevant  operators in the EFT of gravity and higher-spin is identical for all irrelevant operators, hence preventing the presence of high-spin operators at a low scale would necessarily suppress all higher order operators, including the dim-8 ones.
Moreover, even switching off the dim-6 operators by hand, the EFT of gravity with solely the specific dim-8 operator considered in Ref.~\cite{Sennett:2019bpc} is in conflict with causality for $\Lambda < 10^{-11} {\rm eV}$ \cite{deRham:2021bll}, which implies that current GW observations are unable to probe EFTs with vanishing dim-6 operators.

Our analysis demonstrates that the gravitational EFT provides a general and powerful framework to search for beyond GR signatures in inspiralling GWs. While previous studies \cite{Tsukada:2018mbp,Palomba:2019vxe, Sun:2019mqb,Tsukada:2020lgt, KAGRA:2021tse,Yuan:2022bem,Zhang:2021mks,Maselli:2021men} impose constraints on spin-0 and spin-1 particles, our analysis provides the first constraints on higher-spin particles. Comparing to the parameterized post-Newtonian and post-Einstein frameworks, the EFT framework reveals explicit connection between the observational signals and the fundamental physics. The EFT framework can also be used to investigate beyond GR signals in other GW observations, such as extreme-mass-ratio-inspirals \cite{Liu:2023ip} and black hole spectroscopy \cite{Cardoso:2018ptl, deRham:2020ejn, Cano:2021myl}.\\

\begin{acknowledgments}
We thank Claudia de Rham for discussions and for comments on the manuscript.~Y.-S. P. is supported by the NSFC, No.12075246 and the Fundamental Research Funds for the Central Universities. J. Z. is supported by the scientific research starting grants from University of Chinese Academy of Sciences (grant No.~118900M061), the Fundamental Research Funds for the Central Universities (grant No.~E2EG6602X2 and grant No.~E2ET0209X2) and the National Natural Science Foundation of China (NSFC) under grant No.~12147103.
\end{acknowledgments}

\bibliography{references}


\end{document}